\documentclass[prd,twocolumn,,showpacs,showkeys,amsmath,amssymb]{revtex4}

\usepackage{times}
\usepackage{graphicx}

\newcommand{\beq}{\begin{equation}}
\newcommand{\eeq}{\end{equation}}

\newcommand{\ba}{\begin{eqnarray}}
\newcommand{\ea}{\end{eqnarray}}

\def\L{\Lambda}

\def\e{\epsilon}
\def\t{\theta}

\def\ve{\varepsilon}
\def\vt{\vartheta}

\def\gs{\mathrel{\lower0.6ex\hbox{$\buildrel {\textstyle >}\over{\scriptstyle \sim}$}}}
\def\ls{\mathrel{\lower0.6ex\hbox{$\buildrel {\textstyle <}\over{\scriptstyle \sim}$}}}

\begin{document}

\title{On the influence of the cosmological constant on gravitational lensing in small systems}

\author{Mauro Sereno}
\email{sereno@physik.unizh.ch}

\affiliation{
Institut f\"{u}r Theoretische Physik, Universit\"{a}t Z\"{u}rich,
Winterthurerstrasse 190, CH-8057 Z\"{u}rich, Switzerland.
}

\date{January 11, 2008}

\begin{abstract}
The cosmological constant $\L$ affects gravitational lensing phenomena. The contribution of $\L$ to the observable angular positions of multiple images and to their amplification and time delay is here computed through a study in the weak deflection limit of the equations of motion in the Schwarzschild-de Sitter metric. Due to $\L$ the unresolved images are slightly demagnified, the radius of the Einstein ring decreases and the time delay increases. The effect is however negligible for near lenses. In the case of null cosmological constant, we provide some updated results on lensing by a Schwarzschild black hole.
\end{abstract}


\maketitle

\section{Introduction}

The interpretation of the cosmological constant $\L$ is a very fascinating and traditional topic in theoretical physics. On the observational side, large scale structure observations have made a strong case for $\L$ as a possible choice for dark energy. In fact, a very small value of $\L  \sim 10^{-52}\mathrm{m}^{-2}$, together with dark matter, can provide a suitable framework for observational cosmology \cite{pe+ra03}.

Since the cosmological constant should take part in all kinds of gravitational phenomena, investigations have been performed on very different scale-lengths. Despite no convincing method for constraining $\Lambda$ in an Earth's laboratory has been proposed \cite{je+st05}, local astronomical phenomena seem to be more promising. The cosmological constant can influence the motion of massive bodies  \cite{isl83,wri98,ker+al03}. Perihelion precession of solar system planets together with other solar and stellar tests has been considered to put an upper bound of $\L \ls 10^{-42}\mathrm{m}^{-2}$ \cite[and references therein]{wri98,je+se05,ior06,se+je06}. The cosmological constant  also affects the gravitational equilibrium of large astrophysical structures \cite{ba+no05,bal+al06,bam07} and produces a lower velocity dispersion around the Hubble flow on the scale of the Local Volume \cite{tee+al05}.

Recently, \citet{ri+is07} discussed how the cosmological constant takes part in gravitational lensing. Taking into account $\L$ through the Schwarzschild-de Sitter (SdS) metric, they showed that even if the exact differential equation for a light path in the coordinate space can be written in a form that does not involve $\L$ \cite{isl83}, the cosmological constant contributes to the bending of light through the metric itself, which determines the actual observations that can be made on the orbit equation. In fact, one must consider not only the null geodesic equation but also the process of measurement \cite{bak+al07,lak07}. 

Following this correction of the long-standing misconception that $\L$ does not affect the observed deflection angle, in this paper I further investigate the effect of the cosmological constant in gravitational lensing observations in near systems. The weak deflection limit considered throughout the paper allows to have a clear insight on the effect of $\L$ but it is to be remarked that a gravitational lens equation without approximations can be written in generic spherically symmetric and static spacetimes \cite{per04}. In this paper, the lens equation is derived from the lightlike null geodesics of the SdS metric. Results concerning observable quantities are expressed in terms of the invariants of the light ray, avoiding ambiguities connected to coordinate-dependent quantities \cite{ke+pe05,bo+wi03}.

The paper is organized as follows. In Sec.~\ref{sec:geod}, the null orbits are solved in the weak deflection limit. In Sec.~\ref{sec:lens} the lens equation is first written in terms of the observed image position angle and then solved with a perturbation method. Image amplification and time delay are discussed in Sec.~\ref{sec:magn} and~\ref{sec:time}, respectively. Some quantitative estimates of the effect of $\L$ are illustrated in Sec.~\ref{sec:loca} whereas Sec.~\ref{sec:conc} is devoted to some final considerations.

\section{Geodesic equation}
\label{sec:geod}

The role of $\L$ in gravitational lensing can be considered in the framework of the spherically symmetric Schwarzschild vacuum solution with a cosmological constant, also known as Schwarzschild-de~Sitter (SdS) or Kottler space-time \cite{adl+al65},
\beq
\label{sds1}
ds^2=  f_\L (r) dt^2- \frac{dr^2}{f_\L (r)} -r^2 \left( d \theta^2- \sin^2 \theta d\phi^2 \right),
\eeq
where
\beq
\label{sds2} 
f_\L (r) \equiv  \left( 1- \frac{2 m}{r}   -\frac{\L r^2}{3} \right) ,
\eeq
and $m$ is the black hole mass. We are using units $G=c=1$. A coordinate singularity occurs at large radii. For $m=0$, the de Sitter horizon occurs at $r_\L \equiv \sqrt{3/\L}$. Due to spherical symmetry, photon trajectories can be conveniently restricted to the central $\t= \pi/2$ plane. We consider the standard framework of gravitational lensing in the weak deflection limit, where the source of radiation and the observer are remote from the lens.  Lensing in a static, spherically symmetric metric is usually investigated considering asymptotically flat spacetimes  \cite{ke+pe05} but, here, we have to consider both observer and source in a region of spacetime which is well inside the outer horizon. In such a region the intrinsic geometry of the 2-metric of the equatorial plane $\t = \pi/2$ undergoes a transition from a nearly Flamm paraboloid of revolution in the inner region, as typical in Schwarzschild metric, to a spherical geometry of radius $r_\L$ in the very outer and nearly de Sitter spacetime \cite{rin06,ri+is07}. Since the observer lies in this curved transition region of spacetime, even if the null geodesics are formally indistinguishable from the $\L=0$ case in the coordinate space, the observable quantities will be affected by the cosmological constant \cite{ri+is07}. 

In the following analysis the observer and the emitter are taken to be static. The observer coordinates are denoted $\{r_\mathrm{o}, \phi_\mathrm{o} =0 \}$, where $\phi_\mathrm{o}$ has been fixed without loss of generality. The source coordinates are denoted as $\{r_\mathrm{s}, \phi_\mathrm{s} \}$. The orbital equation for a light ray from the source to the observer can then be written in terms of the first integral of motion $b( \equiv \dot{\phi}r^2)$ as
\beq
\label{geo1}
\phi_\mathrm{s} = \pm \int \frac{dr}{ r^2} \left[ \frac{1}{b^2 }+\frac{1}{r_\L^2} - \frac{1}{r^2}  + \frac{2 m}{r^3} \right]^{-1/2} ,
\eeq
where the sign of the integral is adhered to the sign of $dr$ and changes at the inversion points in the $r$-motion. A dot denotes derivation with respect to an affine parameter. Along its path from the source to the observer, the photon passes by the black hole at a minimum distance $r_\mathrm{min}$ which is much larger than the gravitational radius. In the weak deflection limit, this closest approach is the only turning point in the $r$-motion. Defining a new constant $b_\L$ such that $1/b_\L^2 = \left(1/b^2+1/r_\L^2 \right)$, we can see that the geodesics are formally identical to those in a Schwarzschild spacetime without cosmological constant. This can be seen even more clearly taking the second derivative $d^2 r/d\phi^2$, which eliminates $\L$ from the equation. Equation~(\ref{geo1}) can be solved in terms of elliptical functions \cite{lak07} and exact analytical results can be obtained even considering a  spinning black hole \cite{kra05}. In an asymptotically flat spacetime, $b$ can be viewed as the impact parameter.

Even if the equations of motion for either a massive test particle or a photon can be solved exactly \cite{kra05,lak07}, expressions are quite involved, so that lensing observables are more conveniently derived treating through a perturbation approach. A fundamental assumption in the weak deflection limit is that the point of closest approach lies well outside the gravitational radius, i.e. $m/b \equiv \e _\mathrm{m} \ll 1$. The observer and the source lie very far from the lens. It can be shown that $b/r_\mathrm{o} \sim b/r_\mathrm{s} \sim \epsilon_\mathrm{m}$ \cite{ke+pe05}. Furthermore, we assume that the system is embedded in a region well inside the outer horizon, $r_\mathrm{o}, r_\mathrm{s}  \ll r_\L$.  In what follows, we will expand quantities of interest according to the expansion parameters $\e_\mathrm{m}$ and $\e_\L \equiv r_\mathrm{o}/r_\L$ but, for the sake of brevity, we will produce our results up to a given formal order in $\epsilon$, collecting terms coming from any combination of the two expansion parameters \citep{se+de06}. 

The light ray minimum radial distance $r_\mathrm{min}$ to the lens is determined by the equation $r^2=b^2f_\Lambda (r)$, whose exact solution is known analytically \cite{ri+is07}.  Expanding the solution in the weak deflection limit as a power series in $\epsilon$ we find
\beq
\label{geo2}
r_\mathrm{min} \simeq b  \left\{
1 - \frac{m}{b}   -    \frac{3 m^2}{2 b^2}  -   \frac{4 m^3}{b^3}  - \frac{105 m^4}{8 b^4}   -\frac{b^2}{2r_\L^2} 
 \right\} .
\eeq
An expression for the minimum approach including ${\cal{O}}(\epsilon^4)$-terms for the Kerr metric can be found in \cite{se+de06}. In the case of null cosmological constant, equation~(\ref{geo2}) agrees with the result in \cite{ke+pe05}. 

The integral in Eq.~(\ref{geo1}) can be solved approximately under the assumptions discussed above and following standard methods and procedures \cite{ke+pe05,se+de06}. We get
\ba
\phi_\mathrm{s} &  = & -\pi -\frac{4 m}{b} + b \left(\frac{1}{r_\mathrm{s}}+\frac{1}{r_\mathrm{o}}\right)  
-\frac{15 m^2 \pi }{4 b^2} 
-\frac{128 m^3}{3 b^3}   \nonumber \\
& + &  \frac{b^3}{6}  \left(\frac{1}{r_\mathrm{s}^3}+\frac{1}{r_\mathrm{o}^3}\right)  - \frac{3465 m^4 \pi }{64 b^4}
-\frac{3584 m^5}{5 b^5} 
- \frac{2 m b}{r_{\Lambda }^2}        \nonumber \\
 & - &   \frac{ m b^3}{4} \left(\frac{1}{r_\mathrm{s}^4}+\frac{1}{r_\mathrm{o}^4}\right)+  \frac{3 b^5}{40} \left(\frac{1}{r_\mathrm{s}^5}+\frac{1}{r_\mathrm{o}^5}\right) \nonumber  \\
 & - &
 \frac{b^3}{ 2r_{\Lambda }^2}  \left( \frac{1}{r_\mathrm{s}}+\frac{1}{r_\mathrm{o}}\right)
+  {\cal{O}}(\e^6).   \label{geo3}
\ea
The cosmological constant contributes to the geodesic equation through terms of order of  ${\cal{O}}(\e^5)$. The term $2 b m/r_\L^2$, where neither the source or the observer radial position enters, can be considered as local. We are assuming the parameter $b$ to be positive.

\section{Lens equation}
\label{sec:lens}

The lens equation is a mapping relating the position of the source and the observed position of its images. It is usually given in terms of the apparent angular position of the image in the sky, i.e. the angle $\vt$ between the tangent to the photon trajectory at the observer and the radial direction to the black hole as measured in the locally flat observer's frame. In terms of the tetrad components of the four momentum $P$, $\cos \vt = P^{[r]}/ P^{[t]}$. For the SdS metric,
\beq
\label{lens1}
\sin \vt =\sqrt{ f_\L (r_\mathrm{o}) }\frac{b}{r_\mathrm{o}}.
\eeq
The angle $\vt$ is then strictly linked to the constant of motion. For small angles,
\beq
\label{lens2}
\vt \simeq \frac{b}{r_\mathrm{o}} + \frac{b^3}{6 r_\mathrm{o}^3} \left[ 1-\frac{6 m r_\mathrm{o}}{b^2} -\frac{3 r_\mathrm{o}^4}{b^2 r_\L^2}  \right].
\eeq
The repulsive gravitational effect of $\L$ counteracts the attraction of the central mass $m$. Then, light paths seem to be less deflected: once $b$ is fixed, in presence of a non null, positive cosmological constant $\vt$ is smaller than the angle observed when $\L =0$. Due to the presence of $\L$, the relation between $b$ and the observed angle $\vt$ changes by a term of order ${\cal{O}}(\e^3)$, two orders of magnitude higher than the contribution of $\L$ to the variation of the coordinate azimuthal angle, see Eq.~(\ref{geo3}). The relation between the observed angle and the constant of motion determines the extent to which $\L$ affects the lensing observables. In the following resolution of the lens equation, calculations will be then performed up to and including terms of order of ${\cal{O}}(\e^3)$. 

Once we use angular coordinates for the image positions instead on the invariants of motion, it can be appropriate to introduce a series expansion parameter in the weak deflection limit based on the angular Einstein ring defined through radial distances \cite{se+de06}, 
\beq
\label{lens3}
\vt_\mathrm{E} \equiv \sqrt{4 m \frac{r_\mathrm{s}}{r_\mathrm{o}(r_\mathrm{o}+r_\mathrm{s})}} ;
\eeq
the expansion parameter $\ve_\mathrm{E}$ is then defined as $\varepsilon_\mathrm{E}  \equiv  \vt_\mathrm{E}/4 D$ \cite{ke+pe05,se+de06} where $D \equiv r_\mathrm{s}/(r_\mathrm{o}+r_\mathrm{s})$. In a way similar to the case of the geodesic equation, let us perform the expansion in terms of two parameters, $\varepsilon_\mathrm{E} $ and $\e_\L$. Mixed terms are then collected through a given formal order in the parameter $\ve$. The parameters $\varepsilon_\mathrm{E} $ and $\e_\L$ will be written in terms of $\ve$ through the relations $\ve_\mathrm{E} = \ve$  and $\e_\L = \ve/ r_{\L \ve}$, respectively. The parameter $r_{\L \ve}$, defined in the second one of the above equations as the ratio $\ve_\mathrm{E}/ \e_\L$, is usually $\gg 1$ for near astrophysical systems, see Sec.~\ref{sec:loca}.

It is customary in lensing studies to write the source position in terms of the angle $B$ at which the source would be seen in absence of the lens, i.e for $m=0$. In analogy with Eq.~(\ref{lens1}), $B$ is then given by $\sin B =\sqrt{ 1-(r_\mathrm{o}/r_\L)^2 }b_\mathrm{s}/r_0$ with $b_\mathrm{s}$ being a fictitious constant of motion which solves the geodesic motion in Eq.~(\ref{geo1}) for the actual source and observer coordinates but for $m=0$. The azimuthal source coordinate, $\phi_\mathrm{s}$, can then be expressed in terms of $B$ plugging the ``unlensed'' constant $b_\mathrm{s}$ in Eq.~(\ref{geo3}). The lens equation in the form
\beq
\label{lens3b}
{\cal{F}} (B, \vt; m, \L) = 0,
\eeq
is finally obtained by first writing $\phi_\mathrm{s}$ as a function of either $\vt$ or $B$ and then equating the two expressions,
$$
\phi_\mathrm{s}(\vt; m, \L) = \phi_\mathrm{s} (B; m=0, \L) .
$$ 
We will consider source positions $B \geq 0$. At the lowest order, $B \simeq D (\phi_\mathrm{s} + \pi)$.

The lens equation can be solved term by term.  We assume that the solution can be written as a series in $\ve$,
\beq
\label{lens4}
\vt  =  \vt_\mathrm{E} \left\{ \t_0 +\t_1 \ve + \t_2 \ve^2 +{\cal{O}}(\ve^3) \right\}; \nonumber
\eeq
The source position $B$ can be rescaled as $\beta = B/\vt_\mathrm{E}$.  At first order, the lens equation takes the standard form
\beq
\label{lens5}
\beta   = \t_0 -\frac{1}{\theta _0}, 
\eeq
with the usual pair of solutions
\beq
\label{lens6}
\theta _0 ^{\pm}  =  \frac{1}{2} \left(1 \pm \sqrt{1 + \frac{4}{\beta ^2}} \right) \beta .  \nonumber
\eeq
The next order correction is
\beq
\label{lens7}
\theta _1  =  \frac{15 \pi }{16 (1+ \theta _0^2)} . \nonumber
\eeq
Up to and including second order corrections, the cosmological constant is ineffective and lensing is pure Schwarzschild. The cosmological constant shows up at the next order, changing the angular positions of the images as seen by the observer,
\begin{widetext}
\beq
\label{lens8}
\theta _2  =  \frac{8}{\theta _0 \left(\theta_0^2+1\right)}
\left[ 
1+\theta _0^2 -\theta _0^4
+D \left(   1-\frac{7 \theta _0^2}{2}+ \frac{5 \theta _0^4}{2}\right)
-D^2 \left(\frac{2 \theta_0^4}{3} - 2 \theta _0^2 + \frac{5}{3}\right)
 \right]
 -\frac{225 \pi ^2}{256}\frac{1+2 \theta _0^2}{\theta _0 \left(\theta_0^2+1\right)^3}
 -\frac{\theta _0}{r_{\Lambda \varepsilon }^2 \left(\theta _0^2+1\right)}  . \nonumber 
\eeq
\end{widetext}
The effect of the cosmological constant on the observed positions $\vt$ of the images is then $ \propto (\ve /r_{\Lambda \varepsilon})^2 = (r_\mathrm{o}/r_\L)^2 $. The effect depends mainly on the radial distance of the observer but it is also sensitive to the source position trough $\t_0$. The contribution of $\L$ to the angular position of the images can then be written as
\beq
 \delta \vt_\L = - \left( \frac{r_\mathrm{o}}{r_\L} \right)^2  \frac{\theta _0}{1+ \theta _0^2} \vt_\mathrm{E} .
\eeq

The angular splitting between the two images reads
\begin{widetext}
\beq
\label{lens9}
\vt^+ -\vt^- = \vt_\mathrm{E} \left\{ \sqrt{\beta ^2+4} 
-\frac{15 \pi  \beta  \varepsilon }{16 \sqrt{\beta ^2+4}}+
\frac{\ve^2 }{\sqrt{\beta ^2+4}}
\left[ 
16 -\frac{225 \pi^2 \left(\beta ^4+6 \beta ^2+6\right)}{256 \left(\beta ^2+4\right)}  +  28 D \beta ^2  -  \frac{8D^2}{3}  \left(2+ 7 \beta ^2 \right) -\frac{2}{r_{\Lambda \varepsilon }^2} 
   \right] 
\right\} . \nonumber
\eeq
\end{widetext}
Let see how the above results compare to \citep{ri+is07}.  \citet{ri+is07} derived the angle $\vt$, see their equation~(12), for the particular configuration $b \sim \sqrt{2 m r_\mathrm{o}}$, which stands for observer and source at the same radial distance $r_\mathrm{o} \sim r_\mathrm{s}$ ($D =1/2$) and $ b /  r_\mathrm{0} \sim \vt \sim \vt_\mathrm{E}$ ($\t_0=1$). In that case, the contribution of the cosmological constant to the observed image position angle, $\delta \vt_\L$, can be rewritten as $-(1/12) \L b^3 / m$, which agrees with the result in \citep{ri+is07}.

Deflection angle in gravitational lensing is usually defined in asymptotically flat spacetimes as the angle between the asymptotic tangents to the light ray at the observer and at the source. Even though the SdS spacetime is not asymptotically flat, we can identify a sort of contribution of the cosmological constant to the deflection by comparing the lens equations either with or without $\L$, 
\beq
\label{lens9a}
D \frac{\hat{\alpha}_\L}{ \vt_\mathrm{E}} =  {\cal{F}} (B, \vt; m, \L) - {\cal{F}} (B, \vt; m, \L = 0) ,
\eeq
with ${\cal{F}}$ normalized in such a way that at first order it takes the form of Eq.~(\ref{lens5}). The difference is of order of $\ve^3$. As usual, the factor $D$ in the left hand side of Eq.~(\ref{lens9}) allows to turn the ``scaled" deflection angle into the ``effective" one. Using the relation in Eq.~(\ref{lens5}), $\hat{\alpha}_\L$ can be written as
\beq
\label{lens10}
\hat{\alpha}_\L = \hat{\alpha}_\mathrm{pN}  \left( \frac{r_\mathrm{o}}{r_\L}\right)^2= - \frac{4 m r_\mathrm{o} \L}{3 \vt} , 
\eeq
where $\hat{\alpha}_\mathrm{pN} \equiv 4 m /(r_\mathrm{o} \vt)$ is the deflection angle at the post-Newtonian order. We have that $\delta \vt_\L / \hat{\alpha}_\L = D \t_0^2/(1+\t_0^2)$. At a typical angle $\vt = \vt_\mathrm{E}$, 
$$
\hat{\alpha}_\L (\vt_\mathrm{E}) = - \frac{\vt_\mathrm{E}}{D} \left( \frac{r_\mathrm{o}}{r_\L}\right)^2 = - \frac{\vt_\mathrm{E}}{D} \frac{ r_\mathrm{o}^2 \L}{3}.
$$ 

The contribution of the cosmological constant to the lens equation can be derived in an alternative and easier way. The approximate lens equation is usually written in terms of the image position angle, $\vt$, the position angle of the source in absence of the lens, $B$, and angular diameter distances as measured in the smooth background \cite{sch+al92},
\beq
\label{lens11}
B = \vt - \frac{D_\mathrm{ds} }{ D_\mathrm{s}} \hat{\alpha}_\mathrm{pN},
\eeq
where  $\hat{\alpha}_\mathrm{pN}$ is the deflection angle at the post-Newtonian order and $D_\mathrm{ds}$ and $D_\mathrm{s}$ are the angular diameter distances from the lens to the source and from the observer to the source, respectively. In the case we have been considering so far, the black hole $m$ is embedded in an otherwise smooth spacetime which can be described by the de Sitter metric, Eqs.~(\ref{sds1},~\ref{sds2}) for $m=0$. Then, the unperturbed deflection angle in terms of angular diameter distance $D_\mathrm{d}$  from the observer to the lens takes the form $\hat{\alpha}_\mathrm{pN} = 4 m /(D_\mathrm{d} \vt)$, whereas the angular diameter distances can be written in terms of radial coordinates as \cite{sch+al92,sch07}
\begin{eqnarray}
D_\mathrm{d} & = & \frac{r_\mathrm{o}}{\sqrt{1-\left( r_\mathrm{o}/r_\L \right)^2}} ,  \\
D_\mathrm{ds} & = & r_\mathrm{s}  ,   \label{lens12}  \\
D_\mathrm{s} & = & \frac{r_\mathrm{o} + r_\mathrm{s}}{\sqrt{1-\left( r_\mathrm{o}/r_\L \right)^2}} . 
\end{eqnarray}
The above distances have been derived considering static source, lens and observer in the background de~Sitter metric. Plugging Eqs.~(\ref{lens12}) in the lens equation Eq.~(\ref{lens11}) we get
\beq
\label{lens15}
B  =  \vt -  \frac{r_\mathrm{o} }{ r_\mathrm{o} +  r_\mathrm{s} } \frac{4 m}{r_\mathrm{o} \vt}  \left\{  1  -  \left( \frac{r_\mathrm{o}}{r_\L}\right)^2 \right\}.
\eeq
The contribution of the cosmological constant to the lens equation in the right-hand side of Eq.~(\ref{lens15}) has the same form of the expression derived considering the geodesic motion, see Eq.~(\ref{lens10}). Since the main contribution of $\L$ to gravitational lensing comes from the relation between the observed angle and the constant of motion, see Eq.~(\ref{lens1}), it is not surprising that such a contribution can be also obtained by taking care of expressing distances as the angular diameter distances of the background metric. In fact, such distances express the relation between proper physical sizes at the emitter and measurable angles subtended at the observer. In other words, up to order $\e^3$, $\L$ affect lensing phenomena only through the curvature of the background spacetime and does not affect the local deflection of light near the lens. On the other side, it is clear that the cosmological constant affect lensing observations.

It is to be remarked that the above derivation based on the lens equation in the approximate form of Eq.~(\ref{lens11}) allows to determine the main contribution of $\L$ to gravitational lensing but, on the other hand, misses both higher order geometrical corrections and the contributions to the light deflection of post-post-Newtonian order or higher \cite{vi+el00,bo+se06}, which must be properly considered by expanding the geodesics equation.

\section{Magnification}
\label{sec:magn}

The ratio between the angular area of the image in the observer sky and the angular area of the source in absence of lensing gives the (signed) amplification of the image,
\beq
\label{magn1}
\mu = \frac{\sin \vt}{\sin B}\frac{d \vt}{d B} .
\eeq
The magnification of the apparent luminosity is given by correcting such a geometrical amplification for the standard redshift factor. The derivative in Eq.~(\ref{magn1}) can be computed through the chain rule by deriving the coordinate position of the source $\phi_\mathrm{s}$  with respect to either $B$ or $\vt$ and then combining the results suitably. After introducing the scaled angular variables, the result can be rearranged as a series in $\ve$,
\beq
\mu = \mu_0 +\mu_1 \ve + \mu_2 \ve^2 + {\cal{O}}(\ve^3) . \nonumber
\eeq
The first coefficients of the above expansion series are like pure Schwarzschild lensing,
\beq
\mu_0  =  \frac{\theta _0^4}{\theta _0^4-1},  \nonumber  
\eeq
and
\beq
 \mu_1  = -\frac{15 \pi  \theta _0^3}{16 \left(\theta _0^2+1\right)^3}.  \nonumber
\eeq
The $\L$ correction shows up at the next order,
\begin{widetext}
\beq
\mu_2 = \frac{8 \theta _0^2}{(1-\theta _0^2) (1+ \theta _0^2)^3}
\left\{
\theta _0^4 \left(4+ 2\theta _0^2 - \frac{675 \pi ^2}{1024 (1+ \theta _0^2)^2} \right)
+ D \theta _0^2 (9- 10 \theta _0^2 - 5 \theta _0^4)
-\frac{D^2}{3} (1 +16 \theta _0^2 -23 \theta _0^4 -12 \theta _0^6 )
+ \frac{\theta _0^2}{4 r_{\Lambda \varepsilon }^2} 
\right\}  . \nonumber
\eeq
\end{widetext}
Let us consider the microlening case when the two images can not be resolved and the observable is the total magnification $\mu_\mathrm{tot} = |\mu^+| + |\mu^-|$. Using the above results, $\mu_\mathrm{tot}$ can be written in terms of the unlensed source position as
\ba
\mu_\mathrm{tot} & \simeq & \frac{\beta ^2+2}{\beta  \sqrt{\beta ^2+4}}-\frac{15 \pi  \varepsilon }{8 \left(\beta ^2+4\right)^{3/2}}  
- \frac{4 \ve^2}{\beta  \left(\beta ^2+4\right)^{3/2}} 
\nonumber \\
& \times &   \left[ 
\frac{1}{r_{\Lambda \varepsilon }^2}   +
4 (6 +6 \beta ^2+ \beta ^4) - \frac{675 \pi ^2}{256 \left(\beta ^2+4\right)} 
\right. \\
& -& 
 \left.
 2 D (12 +30 \beta ^2 + 5 \beta ^4)
+\frac{4D^2 }{3} (18+35 \beta ^2+ 6 \beta ^4 )
\right] . \nonumber
\ea
The contribution of $\L$ to the total magnification is negative so that images are slightly de-amplified.

The cosmological constant is isotropic and does not perturb the spherical symmetry of the lens. The caustic surface is still a line coincident with the optical axis behind the lens. The tangential critical circle corresponding to the point-like caustics is a perturbed Einstein ring with angular radius
\beq
\vt_\mathrm{t} \simeq  \vt_\mathrm{E} \left\{ 1 +\frac{15\pi}{32}\ve + \left( 4  -\frac{4 D^2}{3}  -\frac{675\pi^2}{2048} -\frac{1}{2 r_{\L \ve}^2}  \right) \ve^2 \right\} . \nonumber
\eeq
Due to $\L$ the area of the Einstein ring slightly decreases.

\section{Time delay}
\label{sec:time}

Light rays corresponding to different images have different travel times. To compute the time delay as measured by an observer we have first to compute the coordinate time $t_\mathrm{o}$ when a given ray reaches the observer position and then to translate the difference from coordinate time to proper time. For the SdS metric
\beq
\label{time1}
t_\mathrm{o} =  \pm \int f_\L(r)^{-1} \left( 1-\frac{b^2}{r^2} f_\L(r) \right)^{-1/2} d r,
\eeq 
where the emission time has been fixed at $t_\mathrm{s}=0$ for all the light rays. The overall sign in Eq.~(\ref{time1}) is adhered to $dr$ to give a positive contribution. Differently from the $r$-motion, the travel time can not be expressed in terms of a new constant of motion $b_\L$ that makes the integral in Eq.~(\ref{time1}) formally identical to the expression for the Schwarzschild metric.  As for the geodesic equation, the travel time can be calculated through an expansion in $\e$. We get
\begin{eqnarray}
t_\mathrm{o} &  \simeq  &   r_\mathrm{o}+r_\mathrm{s} +2 m \left( 1 + \log \frac{4 r_\mathrm{o} r_\mathrm{s}}{b^2} \right) 
-\frac{b^2 }{2} \left(\frac{1}{r_\mathrm{s}}+\frac{1}{r_\mathrm{o}}\right)  \nonumber
\\
& + & \frac{r_\mathrm{o}^3+r_\mathrm{s}^3}{3  r_{\Lambda }^2}      -    \frac{15 m^2 \pi }{2 b}  
+  
\frac{64 m^3}{b^2}    -  \frac{b^4}{8} \left(\frac{1}{r_\mathrm{s}^3}+\frac{1}{r_\mathrm{o}^3}\right)  \label{time2}
\\ 
& - &  4 m^2 \left(\frac{1}{r_\mathrm{s}}+\frac{1}{r_\mathrm{o}}\right) + \frac{2 m \left(r_\mathrm{o}^2+r_\mathrm{s}^2\right)}{r_{\Lambda }^2}   + \frac{m^2 \left(r_\mathrm{o}^3+r_\mathrm{s}^3\right)}{2 r_{\Lambda }^2 b^2}  . \nonumber
\end{eqnarray}
Since an observer measures time differences, only terms in the arrival time containing the impact parameter $b$ contribute to the observed time delay, whereas terms depending either only on the radial positions of source and observer or on $m$ and $\L$ do not. Then the term $\sim (r_\mathrm{o}^3+r_\mathrm{s}^3)/(3  r_{\Lambda }^2)$, which is similar to a contribution already derived in \citep{ker+al03}, can not be measured in lensing observations. The measurable time delay is the interval of proper time between the arrivals of the same intrinsic variation in the source luminosity as observed in each of the two images,
\beq
\label{time3}
\Delta \tau = \sqrt{f_\L (r_\mathrm{o})} ( t_\mathrm{o}^{-}-  t_\mathrm{o}^{+} ) .
\eeq
Expanding in $\ve$ and expressing the result in terms of the angular source position in absence of the lens, we get
\begin{widetext}
\ba
\label{time4}
\Delta \tau  &  =  &
2m  
\left\{  
\delta \tau_0+ \frac{45 \pi }{8} \varepsilon  \sqrt{\beta ^2+4}
+
\varepsilon ^2 \left[
\frac{1}{2 r_{\Lambda \varepsilon }^2}
\left(
\frac{ (1+13D-45 D^2+48 D^3-16 D^4) \beta  \sqrt{\beta ^2+4}}{8 (1-D)^3 D} - 4 D  \delta \tau _0 
\right)  
\right. 
\right.  \\
& + & 
\left. \left. 
\frac{4 \beta }{\sqrt{\beta ^2+4}}
\left[
8 +6 \beta ^2 + \beta ^4+\frac{1575 \pi ^2}{1024} (3+ \beta ^2)  +D( 8-10 \beta ^2-3 \beta ^4 )  -\frac{D^2 }{3} (24 -14 \beta ^2 -5 \beta ^4 )
\right]
-4 D \delta \tau _0 
\right]
\right\} \nonumber
\ea
\end{widetext}
where
\beq
\label{time5}
\delta \tau_0= \beta  \sqrt{\beta ^2+4} +2 \log \frac{ \sqrt{\beta ^2+4} +\beta}{\sqrt{\beta ^2+4}-\beta } . \nonumber
\eeq
Differently from the angular position, the correction term to the time delay due to $\L$  shows factors $D$ and $(1-D)$ at the denominator, so that the effect can be enhanced for sources either very far from ($r_\mathrm{s} \gg r_\mathrm{o}$, $D \rightarrow 1$) or very near  to ($r_\mathrm{s} \ll r_\mathrm{o}$, $D \rightarrow 0$) the lens.

\begin{figure}
        \resizebox{\hsize}{!}{\includegraphics{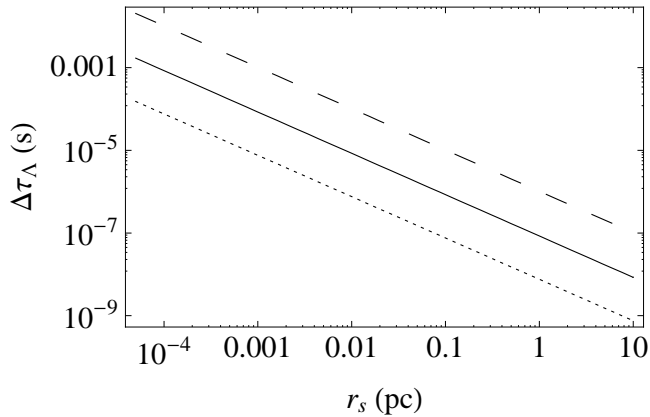}}
        \caption{The contribution to the time delay due to the cosmological constant (in seconds) between the images of a source behind a Sgr~A*-like black hole ($m \sim 3.6\times 10^6 M_\odot$, $ r_\mathrm{o} \sim 7.6~\mathrm{Kpc}$ )  as a function of the source radial distance $r_\mathrm{s}$ (in parsecs). The short-dashed, full and long-dashed lines correspond to source angular positions fixed at $\beta =0.1$, $1$ and $5$, respectively.} 
        \label{fig_delta_tau}
\end{figure}

\section{Near lenses}
\label{sec:loca}

We have seen in the previous sections that the effect of the cosmological constant on lensing observables is really small, being $\sim (r_\mathrm{o}/r_\L)^2$ times smaller than the main post-Newtonian term. It can be nevertheless interesting to give some numbers. A classic test of general relativity consists in measuring the bending of starlight by the Sun. Observations of solar deflection using very long baseline interferometry data allowed to put constraints on deviations from predictions based on the parametrized post-Newtonian formalism at the level of $\ls 0.05 \% $ \cite{leb+al95,sha+al04}. Translating this accuracy into a bound on the cosmological constant, one gets  $\L \ls 10^{-25}\mathrm{m}^{-2}$, nearly 17 orders of magnitude worse than the limits obtained from other solar system tests as precession shift and change in mean motion \cite{se+je06}. 

The supermassive black hole hosted in the radio source Sgr~A* in the Galactic center, with a mass of $\sim 3.6\times 10^6 M_\odot$ and at a distance of $\sim 7.6~\mathrm{kpc}$ from the Earth \citep{eis+al05}, offers another appealing target for testing higher order effects in gravitational lensing with future space- and ground-based experiments \cite{vi+el00,ser04,ke+pe05,se+de06,se+de07}.  For a source $\sim 1~\mathrm{pc}$ behind the black hole, $\L$ induces a variation on the angular position of the images of $\sim 10^{-14}$~arcsec.  Accuracies at the level of $\sim 1~\mu$arcsec, which are within the reach of future missions, are still too low to detect such a tiny effect.

Since multiple images of a single source could be detected behind Sgr~A* in the near future,  prospects for measurements of time delays can deserve some interest. In Fig.~\ref{fig_delta_tau} the time delay due to $\L$ for sources behind Sgr~A* is plotted as a function of the source radial distance, with $r_\mathrm{s}$ spanning the range from $10$~AU to $10$~pc. For sources very near the black hole, the delay can be as large as $10^{-3}$~s.

Since multiple images of a single source could be detected behind Sgr~A* in the near future,  prospects for measurements of time delays can deserve some interest. In Fig.~\ref{fig_delta_tau}, the time delay due to $\L$ for sources behind Sgr~A* is plotted as a function of the source radial distance, with $r_\mathrm{s}$ spanning the range from $10$~AU, slightly smaller than the pericentre of S2 (the observed orbiting star nearest to Sgr~A* \citep{eis+al05}), to $10$~pc, a distance slightly larger than the scale-length of the star cluster in the Galactic center. As you can see from the picture, $\Delta \tau_\L$ increases with the angular separation $\beta$ of the source from the line of sight and decreases with an increasing radial source distance. Due to spherical symmetry, the time delay between the images is null for a source aligned with the line of sight. For sources very near the black hole ($r_\mathrm{s} \ls 10$~AU), the delay can be as large as $10^{-3}$~s. The weak deflection limit is still valid for such a small distance. For $r_\mathrm{s} \ls 10$~AU, $R_\mathrm{Sch} \equiv 2 G m/c^2 \sim 10^{10}~\mathrm{m} < R_\mathrm{E} (\equiv r_\mathrm{o} \vt_\mathrm{E}) \sim 10^{11}~\mathrm{m} < r_\mathrm{s}   \sim 10^{12}~\mathrm{m}$.

Let us finally consider the impact of the cosmological constant on microlensing analyses. A variation $\delta \vt_\mathrm{E}$ in the Einstein radius brings a variation of $2 \delta \vt_\mathrm{E}/\vt_\mathrm{E}$ in the optical depth. Microlensing events have been observed up to the Andromeda galaxy at $\sim 750$~kpc \cite{cal+al02}. Due to $\L$, the optical depth decreases by $\sim 10^{-8}$, which is really negligible.

\section{Conclusions}
\label{sec:conc}

The stagnant theoretical affair between the cosmological constant and the bending of light rays took an hit recently when \citet{ri+is07} pointed out how the study of the orbit equation in the coordinate space is not enough to describe the observations of lensing phenomena. This noteworthy criticism has then stimulated some new interest on the subject \citep{lak07}. In this paper, I have performed an analysis of lensing phenomena in the framework of the SdS metric, which allows a full treatment for systems much smaller than the Hubble radius. I have based my results on a perturbation expansion of the equation of motions, from which I have derived a lens equation accounting for $\L$. The analysis has also showed that the usual argument against $\L$, i.e. that the cosmological constant is dropped out from the exact differential equation for a light path, does not apply to the time delay. It is also to be remarked that the degeneracy between the orbital differential equation in the Schwarzschild metric and that in the SdS spacetime breaks down in presence of a non null angular momentum of the lens.

The argument that $\L$ affects lensing through the metric itself at the observer position is not restricted to the weak deflection limit and applies as well to light rays passing very near to the photon horizon of a black hole. Since SdS null geodesics are formally identical to the Schwarzschild case, the calculation of the deflection angle should be performed as usual but the relation between the constant of motion and the observed angle should be revised. However, since the angular separations of the relativistic images are very small with respect to the splitting of the primary images, in the strong deflection limit it is customary to neglect higher order corrections.

Even though important on a theoretical point of view, the effect of $\L$ on near lenses, such as the Sun, the supermassive black hole in the Galactic center or compact objects in the halo of near galaxies, is quantitatively very small. \citet{ish+al07} tried to extend the result obtained in the framework of the SdS metric to a cosmological scenario where the distances between lens, source and observer are comparable with the Hubble radius. Some caution should be however used in such an extrapolation. The cosmological lens equation is usually derived combining local results on the light deflection, which are based on an asymptotically flat metric, with considerations on global geometry and  angular diameter distances, which are on turn based on the global Friedmann-Lema\^{\i}tre-Robertson-Walker spacetime  in which the system is embedded \cite{sch+al92,sei+al94}. As shown in \citep{ri+is07} and in the present analysis, both based on the SdS metric, the main contribution of $\L$ to lensing observables comes through the value of the metric at the observer position, which lies in a region of spacetime curved by the cosmological constant. In the classical reasoning at the basis of the cosmological lens equation, local effects are related to a small region in the neighborhood of the lens whereas global effects are connected to the large regions of spacetime between source, lens and observer. Then, the main contribution of $\L$ to gravitational lensing should be seen as global in the sense that it is connected to the observer radial distance. This view is also supported by the fact that, as shown in Sec.~\ref{sec:lens}, the effect of $\L$ on the lensing equation can be already considered through the angular diameter distances of the background smooth spacetime, which express global relations.

As far as distances are small with respect to the de Sitter horizon, we can safely apply the expressions obtained in the present analysis and neglect higher order correction connected to the coupling between $\L$ and the black hole mass, but if distances are comparable to the Hubble length then the results should be likely revisited. This will the subject of a future analysis.

A further consideration is that if we are assuming that a constant energy background as the one provided by $\L$ affects lensing, then every other background, such as that provided by dark matter, should have a similar effects. The McVittie metric, which accounts for the presence of a generic cosmological fluid around the central mass and the related expansion of the spacetime, should be used instead of the more specific SdS spacetime and the effect of all the contributions to the cosmological energy budget should be considered even on a small scale \cite{se+je07}.  Then, even though $\L$ changes in some ways the expression for the bending angle, the dark matter, whose uniform distribution counteracts the cosmological constant, should nearly compensate in the opposite direction.

In the case of $\L =0$ the results in this paper updates previous studies for lensing in the Schwarzschild spacetime that were based either on an approximate lens equation, differently from the present analysis which is based on a perturbation analysis of an exact lens equation, or neglected the effect of the metric in the relation between the impact parameter and the observed image position angle. The present study is also relevant to lensing in extended theories of gravity \citep{rug07}, in which the SdS metric provides an exact solution suitable to evaluate the effects of the non linearity of the gravity Lagrangian.

\begin{acknowledgments}
M.S. is supported by the Swiss National Science Foundation and by the Tomalla Foundation.  
\end{acknowledgments}


\begin{thebibliography}{36}
\expandafter\ifx\csname natexlab\endcsname\relax\def\natexlab#1{#1}\fi
\expandafter\ifx\csname bibnamefont\endcsname\relax
  \def\bibnamefont#1{#1}\fi
\expandafter\ifx\csname bibfnamefont\endcsname\relax
  \def\bibfnamefont#1{#1}\fi
\expandafter\ifx\csname citenamefont\endcsname\relax
  \def\citenamefont#1{#1}\fi
\expandafter\ifx\csname url\endcsname\relax
  \def\url#1{\texttt{#1}}\fi
\expandafter\ifx\csname urlprefix\endcsname\relax\def\urlprefix{URL }\fi
\providecommand{\bibinfo}[2]{#2}
\providecommand{\eprint}[2][]{\url{#2}}

\bibitem[{\citenamefont{{Peebles} and {Ratra}}(2003)}]{pe+ra03}
\bibinfo{author}{\bibfnamefont{P.~J.} \bibnamefont{{Peebles}}}
  \bibnamefont{and} \bibinfo{author}{\bibfnamefont{B.}~\bibnamefont{{Ratra}}},
  \bibinfo{journal}{Reviews of Modern Physics} \textbf{\bibinfo{volume}{75}},
  \bibinfo{pages}{559} (\bibinfo{year}{2003}).

\bibitem[{\citenamefont{{Jetzer} and {Straumann}}(2005)}]{je+st05}
\bibinfo{author}{\bibfnamefont{P.}~\bibnamefont{{Jetzer}}} \bibnamefont{and}
  \bibinfo{author}{\bibfnamefont{N.}~\bibnamefont{{Straumann}}},
  \bibinfo{journal}{Physics Letters B} \textbf{\bibinfo{volume}{606}},
  \bibinfo{pages}{77} (\bibinfo{year}{2005}).

\bibitem[{\citenamefont{{Islam}}(1983)}]{isl83}
\bibinfo{author}{\bibfnamefont{J.~N.} \bibnamefont{{Islam}}},
  \bibinfo{journal}{Physics Letters A} \textbf{\bibinfo{volume}{97}},
  \bibinfo{pages}{239} (\bibinfo{year}{1983}).

\bibitem[{\citenamefont{{Wright}}(1998)}]{wri98}
\bibinfo{author}{\bibfnamefont{E.~L.} \bibnamefont{{Wright}}},
  \bibinfo{journal}{astro-ph/9805292}.

\bibitem[{\citenamefont{{Kerr} et~al.}(2003)\citenamefont{{Kerr}, {Hauck}, and
  {Mashhoon}}}]{ker+al03}
\bibinfo{author}{\bibfnamefont{A.~W.} \bibnamefont{{Kerr}}},
  \bibinfo{author}{\bibfnamefont{J.~C.} \bibnamefont{{Hauck}}},
  \bibnamefont{and}
  \bibinfo{author}{\bibfnamefont{B.}~\bibnamefont{{Mashhoon}}},
  \bibinfo{journal}{Classical and Quantum Gravity}
  \textbf{\bibinfo{volume}{20}}, \bibinfo{pages}{2727} (\bibinfo{year}{2003}).

\bibitem[{\citenamefont{{Jetzer} and {Sereno}}(2006)}]{je+se05}
\bibinfo{author}{\bibfnamefont{P.}~\bibnamefont{{Jetzer}}} \bibnamefont{and}
  \bibinfo{author}{\bibfnamefont{M.}~\bibnamefont{{Sereno}}},
  \bibinfo{journal}{Phys Rev. D} \textbf{\bibinfo{volume}{73}},
  \bibinfo{pages}{044015} (\bibinfo{year}{2006}).

\bibitem[{\citenamefont{{Sereno} and {Jetzer}}(2006)}]{se+je06}
\bibinfo{author}{\bibfnamefont{M.}~\bibnamefont{{Sereno}}} \bibnamefont{and}
  \bibinfo{author}{\bibfnamefont{P.}~\bibnamefont{{Jetzer}}},
  \bibinfo{journal}{\prd} \textbf{\bibinfo{volume}{73}},
  \bibinfo{pages}{063004} (\bibinfo{year}{2006}).

\bibitem[{\citenamefont{{Iorio}}(2006)}]{ior06}
\bibinfo{author}{\bibfnamefont{L.}~\bibnamefont{{Iorio}}},
  \bibinfo{journal}{International Journal of Modern Physics D}
  \textbf{\bibinfo{volume}{15}}, \bibinfo{pages}{473} (\bibinfo{year}{2006}).

\bibitem[{\citenamefont{{Balaguera-Antol{\'{\i}}nez} and
  {Nowakowski}}(2005)}]{ba+no05}
\bibinfo{author}{\bibfnamefont{A.}~\bibnamefont{{Balaguera-Antol{\'{\i}}nez}}}
  \bibnamefont{and}
  \bibinfo{author}{\bibfnamefont{M.}~\bibnamefont{{Nowakowski}}},
  \bibinfo{journal}{Astron. Astrophys.} \textbf{\bibinfo{volume}{441}},
  \bibinfo{pages}{23} (\bibinfo{year}{2005}).

\bibitem[{\citenamefont{{Balaguera-Antol{\'{\i}}nez}
  et~al.}(2006)\citenamefont{{Balaguera-Antol{\'{\i}}nez}, {B{\"o}hmer}, and
  {Nowakowski}}}]{bal+al06}
\bibinfo{author}{\bibfnamefont{A.}~\bibnamefont{{Balaguera-Antol{\'{\i}}nez}}},
  \bibinfo{author}{\bibfnamefont{C.~G.} \bibnamefont{{B{\"o}hmer}}},
  \bibnamefont{and}
  \bibinfo{author}{\bibfnamefont{M.}~\bibnamefont{{Nowakowski}}},
  \bibinfo{journal}{Classical Quantum Gravity} \textbf{\bibinfo{volume}{23}},
  \bibinfo{pages}{485} (\bibinfo{year}{2006}).

\bibitem[{\citenamefont{{Bambi}}(2007)}]{bam07}
\bibinfo{author}{\bibfnamefont{C.}~\bibnamefont{{Bambi}}},
  \bibinfo{journal}{\prd} \textbf{\bibinfo{volume}{75}},
  \bibinfo{pages}{083003} (\bibinfo{year}{2007}).

\bibitem[{\citenamefont{{Teerikorpi} et~al.}(2005)\citenamefont{{Teerikorpi},
  {Chernin}, and {Baryshev}}}]{tee+al05}
\bibinfo{author}{\bibfnamefont{P.}~\bibnamefont{{Teerikorpi}}},
  \bibinfo{author}{\bibfnamefont{A.~D.} \bibnamefont{{Chernin}}},
  \bibnamefont{and} \bibinfo{author}{\bibfnamefont{Y.~V.}
  \bibnamefont{{Baryshev}}}, \bibinfo{journal}{Astron. Astrophys.}
  \textbf{\bibinfo{volume}{440}}, \bibinfo{pages}{791} (\bibinfo{year}{2005}).

\bibitem[{\citenamefont{{Rindler} and {Ishak}}(2007)}]{ri+is07}
\bibinfo{author}{\bibfnamefont{W.}~\bibnamefont{{Rindler}}} \bibnamefont{and}
  \bibinfo{author}{\bibfnamefont{M.}~\bibnamefont{{Ishak}}},
  \bibinfo{journal}{\prd} \textbf{\bibinfo{volume}{76}},
  \bibinfo{pages}{043006} (\bibinfo{year}{2007}).

\bibitem[{\citenamefont{{Lake}}(2007)}]{lak07}
\bibinfo{author}{\bibfnamefont{K.}~\bibnamefont{{Lake}}},
  \bibinfo{journal}{ArXiv: 0711.0673v1}.

\bibitem[{\citenamefont{{Bakala} et~al.}(2007)\citenamefont{{Bakala}, {Cermak},
  {Hledik}, {Stuchlik}, and {Truparova}}}]{bak+al07}
\bibinfo{author}{\bibfnamefont{P.}~\bibnamefont{{Bakala}}},
  \bibinfo{author}{\bibfnamefont{P.}~\bibnamefont{{Cermak}}},
  \bibinfo{author}{\bibfnamefont{S.}~\bibnamefont{{Hledik}}},
  \bibinfo{author}{\bibfnamefont{Z.}~\bibnamefont{{Stuchlik}}},
  \bibnamefont{and}
  \bibinfo{author}{\bibfnamefont{K.}~\bibnamefont{{Truparova}}},
  \bibinfo{journal}{ArXiv: 0709.4274}.

\bibitem[{\citenamefont{{Perlick}}(2004)}]{per04}
\bibinfo{author}{\bibfnamefont{V.}~\bibnamefont{{Perlick}}},
  \bibinfo{journal}{\prd} \textbf{\bibinfo{volume}{69}},
  \bibinfo{pages}{064017} (\bibinfo{year}{2004}).

\bibitem[{\citenamefont{{Keeton} and {Petters}}(2005)}]{ke+pe05}
\bibinfo{author}{\bibfnamefont{C.~R.} \bibnamefont{{Keeton}}} \bibnamefont{and}
  \bibinfo{author}{\bibfnamefont{A.~O.} \bibnamefont{{Petters}}},
  \bibinfo{journal}{\prd} \textbf{\bibinfo{volume}{72}},
  \bibinfo{pages}{104006} (\bibinfo{year}{2005}).

\bibitem[{\citenamefont{{Bodenner} and {Will}}(2003)}]{bo+wi03}
\bibinfo{author}{\bibfnamefont{J.}~\bibnamefont{{Bodenner}}} \bibnamefont{and}
  \bibinfo{author}{\bibfnamefont{C.~M.} \bibnamefont{{Will}}},
  \bibinfo{journal}{Am. J. Phys.} \textbf{\bibinfo{volume}{71}},
  \bibinfo{pages}{770} (\bibinfo{year}{2003}).

\bibitem[{\citenamefont{{Adler} et~al.}(1965)\citenamefont{{Adler}, {Bazin},
  and {Schiffer}}}]{adl+al65}
\bibinfo{author}{\bibfnamefont{R.}~\bibnamefont{{Adler}}},
  \bibinfo{author}{\bibfnamefont{M.}~\bibnamefont{{Bazin}}}, \bibnamefont{and}
  \bibinfo{author}{\bibfnamefont{M.}~\bibnamefont{{Schiffer}}},
  \emph{\bibinfo{title}{{Introduction to general relativity}}}
  (\bibinfo{publisher}{New York: McGraw-Hill}, \bibinfo{year}{1965}).

\bibitem[{\citenamefont{{Rindler}}(2006)}]{rin06}
\bibinfo{author}{\bibfnamefont{W.}~\bibnamefont{{Rindler}}},
  \emph{\bibinfo{title}{{General Relativity}}} (\bibinfo{publisher}{University
  of Chicago Press, Chicago, 2nd ed.}, \bibinfo{year}{2006}).

\bibitem[{\citenamefont{{Kraniotis}}(2005)}]{kra05}
\bibinfo{author}{\bibfnamefont{G.~V.} \bibnamefont{{Kraniotis}}},
  \bibinfo{journal}{Classical and Quantum Gravity}
  \textbf{\bibinfo{volume}{22}}, \bibinfo{pages}{4391} (\bibinfo{year}{2005}).

\bibitem[{\citenamefont{{Sereno} and {De Luca}}(2006)}]{se+de06}
\bibinfo{author}{\bibfnamefont{M.}~\bibnamefont{{Sereno}}} \bibnamefont{and}
  \bibinfo{author}{\bibfnamefont{F.}~\bibnamefont{{De Luca}}},
  \bibinfo{journal}{\prd} \textbf{\bibinfo{volume}{74}},
  \bibinfo{pages}{123009} (\bibinfo{year}{2006}).

\bibitem[{\citenamefont{{Schneider} et~al.}(1992)\citenamefont{{Schneider},
  {Ehlers}, and {Falco}}}]{sch+al92}
\bibinfo{author}{\bibfnamefont{P.}~\bibnamefont{{Schneider}}},
  \bibinfo{author}{\bibfnamefont{J.}~\bibnamefont{{Ehlers}}}, \bibnamefont{and}
  \bibinfo{author}{\bibfnamefont{E.~E.} \bibnamefont{{Falco}}},
  \emph{\bibinfo{title}{{Gravitational Lenses}}}
  (\bibinfo{publisher}{Springer-Verlag Berlin}, \bibinfo{year}{1992}).

\bibitem[{\citenamefont{{Schucker}}(2007)}]{sch07}
\bibinfo{author}{\bibfnamefont{T.}~\bibnamefont{{Schucker}}},
  \bibinfo{journal}{ArXiv: 0712.1559}.

\bibitem[{\citenamefont{{Virbhadra} and {Ellis}}(2000)}]{vi+el00}
\bibinfo{author}{\bibfnamefont{K.~S.} \bibnamefont{{Virbhadra}}}
  \bibnamefont{and} \bibinfo{author}{\bibfnamefont{G.~F.~R.}
  \bibnamefont{{Ellis}}}, \bibinfo{journal}{\prd}
  \textbf{\bibinfo{volume}{62}}, \bibinfo{pages}{084003}
  (\bibinfo{year}{2000}).

\bibitem[{\citenamefont{{Bozza} and {Sereno}}(2006)}]{bo+se06}
\bibinfo{author}{\bibfnamefont{V.}~\bibnamefont{{Bozza}}} \bibnamefont{and}
  \bibinfo{author}{\bibfnamefont{M.}~\bibnamefont{{Sereno}}},
  \bibinfo{journal}{\prd} \textbf{\bibinfo{volume}{73}},
  \bibinfo{pages}{103004} (\bibinfo{year}{2006}).

\bibitem[{\citenamefont{{Lebach} et~al.}(1995)\citenamefont{{Lebach}, {Corey},
  {Shapiro}, {Ratner}, {Webber}, {Rogers}, {Davis}, and {Herring}}}]{leb+al95}
\bibinfo{author}{\bibfnamefont{D.~E.} \bibnamefont{{Lebach}}},
  \bibinfo{author}{\bibfnamefont{B.~E.} \bibnamefont{{Corey}}},
  \bibinfo{author}{\bibfnamefont{I.~I.} \bibnamefont{{Shapiro}}},
  \bibinfo{author}{\bibfnamefont{M.~I.} \bibnamefont{{Ratner}}},
  \bibinfo{author}{\bibfnamefont{J.~C.} \bibnamefont{{Webber}}},
  \bibinfo{author}{\bibfnamefont{A.~E.~E.} \bibnamefont{{Rogers}}},
  \bibinfo{author}{\bibfnamefont{J.~L.} \bibnamefont{{Davis}}},
  \bibnamefont{and} \bibinfo{author}{\bibfnamefont{T.~A.}
  \bibnamefont{{Herring}}}, \bibinfo{journal}{Physical Review Letters}
  \textbf{\bibinfo{volume}{75}}, \bibinfo{pages}{1439} (\bibinfo{year}{1995}).

\bibitem[{\citenamefont{{Shapiro} et~al.}(2004)\citenamefont{{Shapiro},
  {Davis}, {Lebach}, and {Gregory}}}]{sha+al04}
\bibinfo{author}{\bibfnamefont{S.~S.} \bibnamefont{{Shapiro}}},
  \bibinfo{author}{\bibfnamefont{J.~L.} \bibnamefont{{Davis}}},
  \bibinfo{author}{\bibfnamefont{D.~E.} \bibnamefont{{Lebach}}},
  \bibnamefont{and} \bibinfo{author}{\bibfnamefont{J.~S.}
  \bibnamefont{{Gregory}}}, \bibinfo{journal}{Physical Review Letters}
  \textbf{\bibinfo{volume}{92}}, \bibinfo{pages}{121101}
  (\bibinfo{year}{2004}).

\bibitem[{\citenamefont{{Eisenhauer} et~al.}(2005)\citenamefont{{Eisenhauer},
  {Genzel}, {Alexander}, {Abuter}, {Paumard}, {Ott}, {Gilbert}, {Gillessen},
  {Horrobin}, {Trippe} et~al.}}]{eis+al05}
\bibinfo{author}{\bibfnamefont{F.}~\bibnamefont{{Eisenhauer}}},
  \bibinfo{author}{\bibfnamefont{R.}~\bibnamefont{{Genzel}}},
  \bibinfo{author}{\bibfnamefont{T.}~\bibnamefont{{Alexander}}},
  \bibinfo{author}{\bibfnamefont{R.}~\bibnamefont{{Abuter}}},
  \bibinfo{author}{\bibfnamefont{T.}~\bibnamefont{{Paumard}}},
  \bibinfo{author}{\bibfnamefont{T.}~\bibnamefont{{Ott}}},
  \bibinfo{author}{\bibfnamefont{A.}~\bibnamefont{{Gilbert}}},
  \bibinfo{author}{\bibfnamefont{S.}~\bibnamefont{{Gillessen}}},
  \bibinfo{author}{\bibfnamefont{M.}~\bibnamefont{{Horrobin}}},
  \bibinfo{author}{\bibfnamefont{S.}~\bibnamefont{{Trippe}}},
  \bibnamefont{et~al.}, \bibinfo{journal}{\apj} \textbf{\bibinfo{volume}{628}},
  \bibinfo{pages}{246} (\bibinfo{year}{2005}).

\bibitem[{\citenamefont{{Sereno} and {De Luca}}(2007)}]{se+de07}
\bibinfo{author}{\bibfnamefont{M.}~\bibnamefont{{Sereno}}} \bibnamefont{and}
  \bibinfo{author}{\bibfnamefont{F.}~\bibnamefont{{De Luca}}},
  \bibinfo{journal}{ArXiv: 0710.5923}.

\bibitem[{\citenamefont{{Sereno}}(2004)}]{ser04}
\bibinfo{author}{\bibfnamefont{M.}~\bibnamefont{{Sereno}}},
  \bibinfo{journal}{\prd} \textbf{\bibinfo{volume}{69}},
  \bibinfo{pages}{023002} (\bibinfo{year}{2004}).

\bibitem[{\citenamefont{{Calchi Novati} et~al.}(2002)\citenamefont{{Calchi
  Novati}, {Iovane}, {Marino}, {Auri{\`e}re}, {Baillon}, {Bouquet}, {Bozza},
  {Capaccioli}, {Capozziello}, {Cardone} et~al.}}]{cal+al02}
\bibinfo{author}{\bibfnamefont{S.}~\bibnamefont{{Calchi Novati}}},
  \bibinfo{author}{\bibfnamefont{G.}~\bibnamefont{{Iovane}}},
  \bibinfo{author}{\bibfnamefont{A.~A.} \bibnamefont{{Marino}}},
  \bibinfo{author}{\bibfnamefont{M.}~\bibnamefont{{Auri{\`e}re}}},
  \bibinfo{author}{\bibfnamefont{P.}~\bibnamefont{{Baillon}}},
  \bibinfo{author}{\bibfnamefont{A.}~\bibnamefont{{Bouquet}}},
  \bibinfo{author}{\bibfnamefont{V.}~\bibnamefont{{Bozza}}},
  \bibinfo{author}{\bibfnamefont{M.}~\bibnamefont{{Capaccioli}}},
  \bibinfo{author}{\bibfnamefont{S.}~\bibnamefont{{Capozziello}}},
  \bibinfo{author}{\bibfnamefont{V.}~\bibnamefont{{Cardone}}},
  \bibnamefont{et~al.}, \bibinfo{journal}{Astron. Astroph.}
  \textbf{\bibinfo{volume}{381}}, \bibinfo{pages}{848} (\bibinfo{year}{2002}).

\bibitem[{\citenamefont{{Ishak} et~al.}(2007)\citenamefont{{Ishak}, {Rindler},
  {Dossett}, {Moldenhauer}, and {Allison}}}]{ish+al07}
\bibinfo{author}{\bibfnamefont{M.}~\bibnamefont{{Ishak}}},
  \bibinfo{author}{\bibfnamefont{W.}~\bibnamefont{{Rindler}}},
  \bibinfo{author}{\bibfnamefont{J.}~\bibnamefont{{Dossett}}},
  \bibinfo{author}{\bibfnamefont{J.}~\bibnamefont{{Moldenhauer}}},
  \bibnamefont{and}
  \bibinfo{author}{\bibfnamefont{C.}~\bibnamefont{{Allison}}},
  \bibinfo{journal}{ArXiv: 0710.4726v1}.

\bibitem[{\citenamefont{{Seitz} et~al.}(1994)\citenamefont{{Seitz},
  {Schneider}, and {Ehlers}}}]{sei+al94}
\bibinfo{author}{\bibfnamefont{S.}~\bibnamefont{{Seitz}}},
  \bibinfo{author}{\bibfnamefont{P.}~\bibnamefont{{Schneider}}},
  \bibnamefont{and} \bibinfo{author}{\bibfnamefont{J.}~\bibnamefont{{Ehlers}}},
  \bibinfo{journal}{Classical and Quantum Gravity}
  \textbf{\bibinfo{volume}{11}}, \bibinfo{pages}{2345} (\bibinfo{year}{1994}).

\bibitem[{\citenamefont{{Sereno} and {Jetzer}}(2007)}]{se+je07}
\bibinfo{author}{\bibfnamefont{M.}~\bibnamefont{{Sereno}}} \bibnamefont{and}
  \bibinfo{author}{\bibfnamefont{P.}~\bibnamefont{{Jetzer}}},
  \bibinfo{journal}{\prd} \textbf{\bibinfo{volume}{75}},
  \bibinfo{pages}{064031} (\bibinfo{year}{2007}).

\bibitem[{\citenamefont{{Ruggiero}}(2007)}]{rug07}
\bibinfo{author}{\bibfnamefont{M.~L.} \bibnamefont{{Ruggiero}}},
  \bibinfo{journal}{ArXiv: 0712.3218}.

\end{thebibliography}

\end{document}